\documentclass[prl,twocolumn,amssymb,showpacs]{revtex4}

\usepackage{bm}
\usepackage{graphicx}
\usepackage{dcolumn}
\begin{document}

\title{The stationary density matrix of a pumped polariton system}
\author{Carlos Andr\'es Vera$^{(a)}$, Alejandro Cabo$^{(b)}$, and 
 Augusto Gonz\'alez$^{(b)}$}
\affiliation{$^{(a)}$Instituto de F\'{\i}sica, Universidad de Antioquia,
 AA 1226, Medell\'{\i}n, Colombia\\
 $^{(b)}$Instituto de Cibern\'etica, Matem\'atica y 
 F\'{\i}sica, Calle E 309, Vedado, Ciudad Habana, Cuba}

\begin{abstract}
The density matrix, $\rho$, of a model polariton system is obtained
numerically from a master equation which takes account of pumping and
losses. In the stationary limit, the coherences between 
eigenstates of the Hamiltonian are three orders of magnitude smaller
than the occupations, meaning that the stationary density matrix is
approximately diagonal in the energy representation. A weakly distorted 
grand canonical Gibbs distribution fits well the occupations. 
\end{abstract}

\maketitle

The possibility of Bose-Einstein condensation (BEC) of excitonic 
polaritons in microcavities has raised big expectations recently 
\cite{Yamamoto,LeSiDang,otros}. Due to the relatively small lifetimes 
of polaritons (of the order of picoseconds) and still smaller rates for
phonon relaxation in the lower polariton branch \cite{r1}, a very 
important question should be answered concerning whether the observed 
magnitudes come from a system in thermal equilibrium or have a 
dynamical nature.

In paper [\onlinecite{nuestraPRL}], we show that in a decaying system
the enhancement of ground-state occupations can be understood in terms
of the combined effects of polariton-polariton scattering and photon
emission, even if phonon relaxation does not act. This suggests that
the results of the experiment reported in [\onlinecite{Yamamoto}] could
be ascribed to a dynamical effect and not necessarily to BEC in the
polariton system.

On the other hand, in the continuously pumped system, where a stationary
state is reached when pump and losses are equilibrated, it was 
undoubtedly demonstrated that phonon relaxation is not effective in
the lower polariton states \cite{JBloch}. Thus, the question arises 
about what kind of stationary state are we reaching in the experiments
reported in [\onlinecite{LeSiDang}].

In the present paper, we are aimed at giving a partial answer to the 
latter question. We will assume that there are not thermalization
mechanisms, and will compute the density matrix arising from a purely
dynamical equation. We use the same model of polariton system, previously
studied in Refs. [\onlinecite{nuestraPRL,PhysE35(2006)99}], with a
finite number of single-particle states for electrons and holes and a
single photon mode in the microcavity. A term accounting for pumping
is added to the master equation for the density matrix. This master
equation is numerically solved in order to find the stationary density 
matrix. The main result of the paper is the following: in the 
stationary limit the density matrix is approximately diagonal in the 
energy representation and, thus, describes a kind of quasiequilibrium 
which can be fitted to a weakly distorted grand canonical Gibbs 
distribution. The distorted Gibbs distribution can be thought of as 
coming from the maximization of the entropy with an
additional constraint in phase space commuting with the Hamiltonian and
fulfilling the requirement of additivity. This idea was presented in 
Ref. [\onlinecite{Cabo}] to describe
quasi stationary nonequilibrium states and share similarities with the
results of Hamiltonian dynamics simulations \cite{italianos}. 

For completeness, we first recall the main features of the model 
polariton system. The Hamiltonian describing the system is the 
following:

\begin{eqnarray}
H&=& \sum_{i}\left\{T^{(e)}_{i}e^\dagger_i e_i+ T^{(h)}_{\bar i}
h^\dagger_{\bar i} h_{\bar i}\right\} \nonumber\\
&+& \frac{\beta}{2} \sum_{ijkl}\langle ij||kl\rangle~ 
e^\dagger_i e^\dagger_j e_l e_k +\frac{\beta}{2} 
\sum_{\bar i\bar j\bar k\bar l}\langle \bar i\bar j||\bar k\bar l\rangle
~ h^\dagger_{\bar i} h^\dagger_{\bar j} h_{\bar l} h_{\bar k}
\nonumber\\
&-& \beta \sum_{i\bar j k\bar l}\langle i\bar j||k\bar l\rangle~ 
e^\dagger_{i} h^\dagger_{\bar j} h_{\bar l} e_{k} + (E_{gap}+
\Delta)~ a^\dagger a\nonumber\\
&+& g \sum_i \left\{ a^\dagger h_{\bar i}e_i+
a e^\dagger_i h^\dagger_{\bar i}\right\}.
\label{eq1}
\end{eqnarray}

\noindent
We include 10 single-electron and 10 single-hole levels in Eq. (\ref{eq1})
(the first three two-dimensional harmonic-oscillator shells)  .
The single-particle spectrum for electrons and holes is assumed flat:

\begin{equation}
T^{(e)}_{i}=E_{gap}, ~~T^{(h)}_{\bar i}=0.
\end{equation}

\noindent
This means that the dot confinement energy, $\hbar\omega_0$, is assumed 
much smaller than the effective band gap, $E_{gap}$. Our model describes a
relatively small quantum dot strongly interacting with the lowest photon
mode of a thin microcavity. $\beta$ is the characteristic Coulomb energy,
$\beta=e^2/(4\pi\epsilon l_{osc})=e^2/(4\pi\epsilon)
\sqrt{m\omega_0/\hbar}$, where $\epsilon$ is the medium dielectric 
constant, and $l_{osc}=\sqrt{\hbar/(m\omega_0)}$ -- the oscillator length. 
We will take the value, $\beta=$2 meV. $\langle ij||kl\rangle$ are matrix 
elements of the Coulomb interaction between harmonic oscillator states. 
The parameter $\Delta$ gives the detuning of the photon energy 
with respect to the (bare) pair energy, equal to $E_{gap}$, and $g=$3 meV 
is the photon-matter coupling strength. Notice that the photon couples the 
electron state $i$ to the hole state $\bar i$, which differs from the 
latter only in the sign of the angular momentum. This means that
electron-hole pairs are created or annihilated in states with zero angular
momentum. As a result, the Hamiltonian, Eq. (\ref{eq1}), commutes with the  
total angular momentum of the system, $L_{total}$. In what follows, we
consider only $L_{total}=0$ states.

The Hamiltonian, Eq. (\ref{eq1}), preserves the polariton number,

\begin{equation}
N_{pol}=a^\dagger a+\sum_i (h^\dagger_{\bar i}h_{\bar i}+ 
e^\dagger_i e_i)/2.
\end{equation}

\begin{figure}[t]
\begin{center}
\includegraphics[width=.95\linewidth,angle=0]{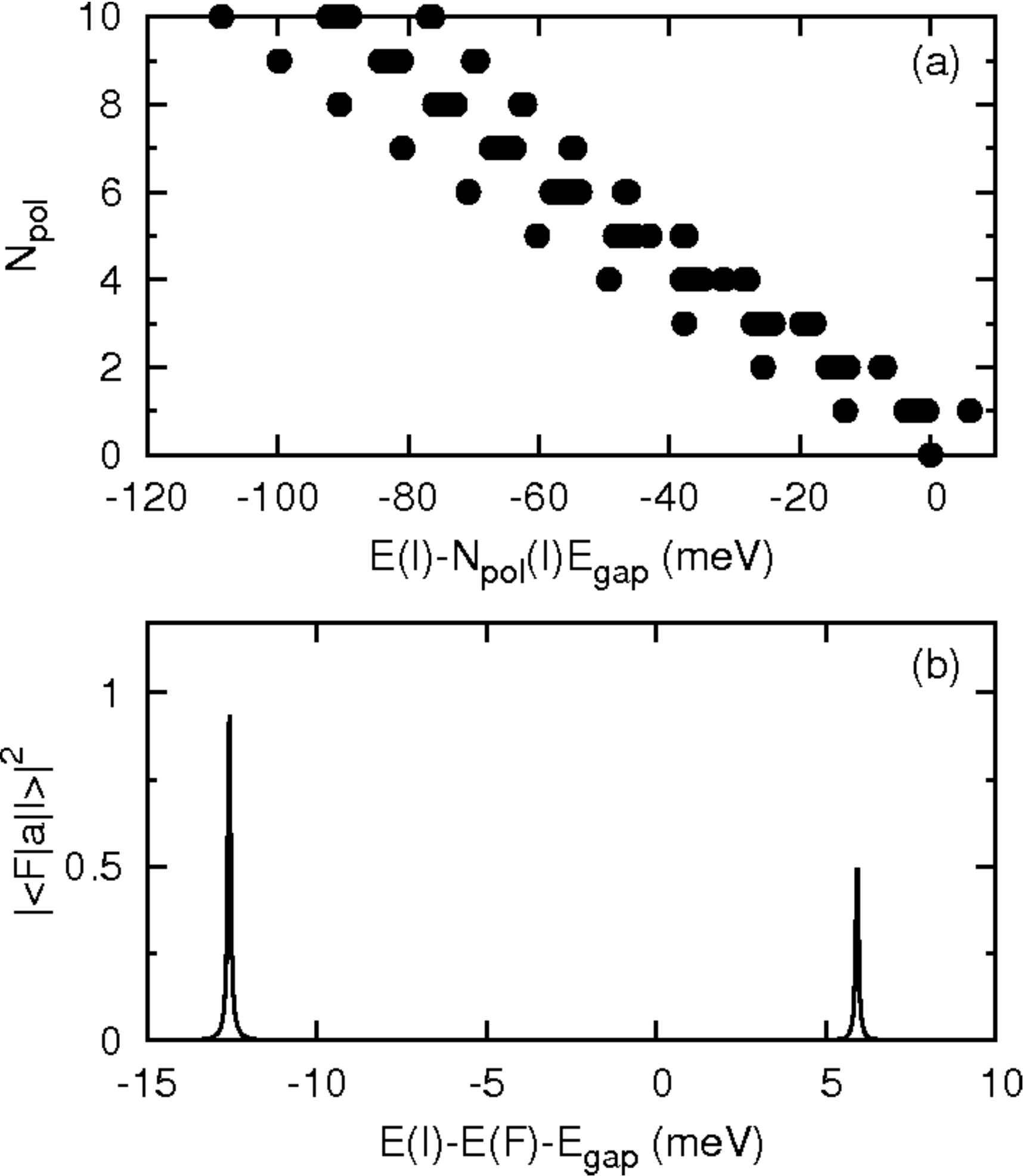}
\caption{\label{fig1} (a) The lowest states in sectors with 
$0\le N_{pol}\le 10$. (b) The matrix elements $|\langle F|a|I\rangle|^2$, 
smeared with a Lorentzian of width $\Gamma=0.1$ meV, vs the energy
difference $E(I)-E(F)-E_{gap}$. The states $|I\rangle$ correspond to
$N_{pol}=2$, whereas $|F\rangle$ is the ground state in the $N_{pol}=1$ 
sector}
\end{center}
\end{figure}

We diagonalize the Hamiltonian in a basis constructed from Slater
determinants for electrons and holes and Fock states of photons. For a 
given polariton number, $N_{pol}$, the
wave functions are looked for as linear combinations:

\begin{equation}
|I\rangle=\sum C_{S_e,S_h,n} |S_e,S_h,n\rangle.
\end{equation}

\noindent
where $S_e$ and $S_h$ are Slater determinants with the same number of
particles, $N_{pairs}$, and $n=N_{pol}-N_{pairs}$. When $N_{pol}=0$
there is only one state, the vacuum. When $N_{pol}=1$ there are 17
states with $L_{total}=0$. One of them is the state with one photon (no
pairs), and the remaining 16 states correspond to matter excitations 
(no photons), that is, all possible combinations of one electron and one 
hole states with total angular momentum equal to zero. As $N_{pol}$ 
increases, the number of eigenstates of $H$ rises, reaching around 18000 
for $N_{pol}=10$. We use Lanczos algorithms \cite{Lanczos} to obtain the 
energies and wavefunctions of the lowest 20 states, which are used to 
write down the master equation for the density matrix. This number of 
states, 20, is chosen on the basis of two reasons. First, we want to keep 
the number of matrix elements, $\rho_{FI}$, between reasonable limits. And, 
second, in a sector with given $N_{pol}$ we shall include all of the 
states, $|I\rangle$, with significant matrix elements, 
$\langle F|a|I\rangle$, between $|I\rangle$ and the ground state in the 
sector with $N_{pol}-1$. They are very important in the dynamics, as it 
will be shown below.

We draw in Fig. \ref{fig1}(a) the set of states used in this work for a
detuning $\Delta=-3$ meV. Notice the approximate linear
dependence of the ground-state energy on $N_{pol}$, and the
energy gap from the ground to the first excited state in each sector, 
roughly proportional to $\sqrt{N_{pol}}~g$. 


A sample of the matrix elements $|\langle F|a|I\rangle|^2$ is
represented in Fig. \ref{fig1}(b). Matrix elements are smeared out with a 
Lorentzian of width $\Gamma=0.1$ meV. The states 
$|I\rangle$ correspond to $N_{pol}=2$, whereas
$|F\rangle$ is the ground state in the $N_{pol}=1$ sector. The lower
(LP) and upper polariton (UP) branches are clearly distinguished. 
Notice that
the latter are included among the 20 states which participate in the
dynamics. This is important because they substantially contribute
to the population of the lowest state in each sector with given
$N_{pol}$.

Now we come to the central point of the paper, the master equation for
the density matrix and its stationary solution. The master equation is
written as \cite{Tejedor,libroOC}:

\begin{eqnarray}
\frac{{\rm d}\rho}{{\rm d}t}&=&-\frac{i}{\hbar} [H,\rho]
+\frac{\kappa}{2} (2 a\rho a^{\dagger}-a^{\dagger} a \rho-
\rho a^{\dagger}a)\nonumber\\
&+&\frac{P}{2} \sum_{I,J}(2\sigma_{IJ}^{\dagger}\rho\sigma_{IJ}-
\sigma_{IJ}\sigma_{IJ}^{\dagger}\rho
-\rho\sigma_{IJ}\sigma_{IJ}^{\dagger}). 
\label{eq5}
\end{eqnarray}

\noindent
The parameter $\kappa$ accounts for photon losses through the cavity 
mirrors ($\hbar\kappa\approx E_{gap}/Q$, where $Q$ is the cavity 
quality factor). In our calculations, we take $\kappa=0.1$ ps$^{-1}$. 
Notice that $\kappa << g$, thus our model system works under the strong
light-matter coupling regime. Other sources of losses such as, for
example, spontaneous exciton decay are much less important and will not
be considered. On the other hand, the parameter $P$ is a pumping rate.
In our modeling, incoherent pumping is supposed to come from highly 
excited excitonic states, which decay towards the lower polariton states
through phonon emission. This is a kind of polariton reservoir. We will
use a sort of homogeneous pumping, with equal probabilities for all
states. To this end, we introduce lowering and rising operators, 
$\sigma_{IJ}|I\rangle=|J\rangle$, 
$\sigma_{IJ}^{\dagger}|J\rangle=|I\rangle$. As we are employing a 
finite number of states (20) in each sector with given $N_{pol}>1$, 
total pumping probabilities are finite. Notice that we are not
including relaxation acting at the ``horizontal'' level in Fig. 
\ref{fig1}(a), i.e. causing the excited polariton states to decay towards
the lowest states with the same $N_{pol}$. Thus, if the lowest polariton
states become occupied in our simulation it is the result of dynamics
and not of relaxation. The absence of phonon thermalization is also the
reason why $L_{total}=0$ states are decoupled from other states with
$L_{total}\ne 0$.

\begin{figure}[t]
\begin{center}
\includegraphics[width=.96\linewidth,angle=0]{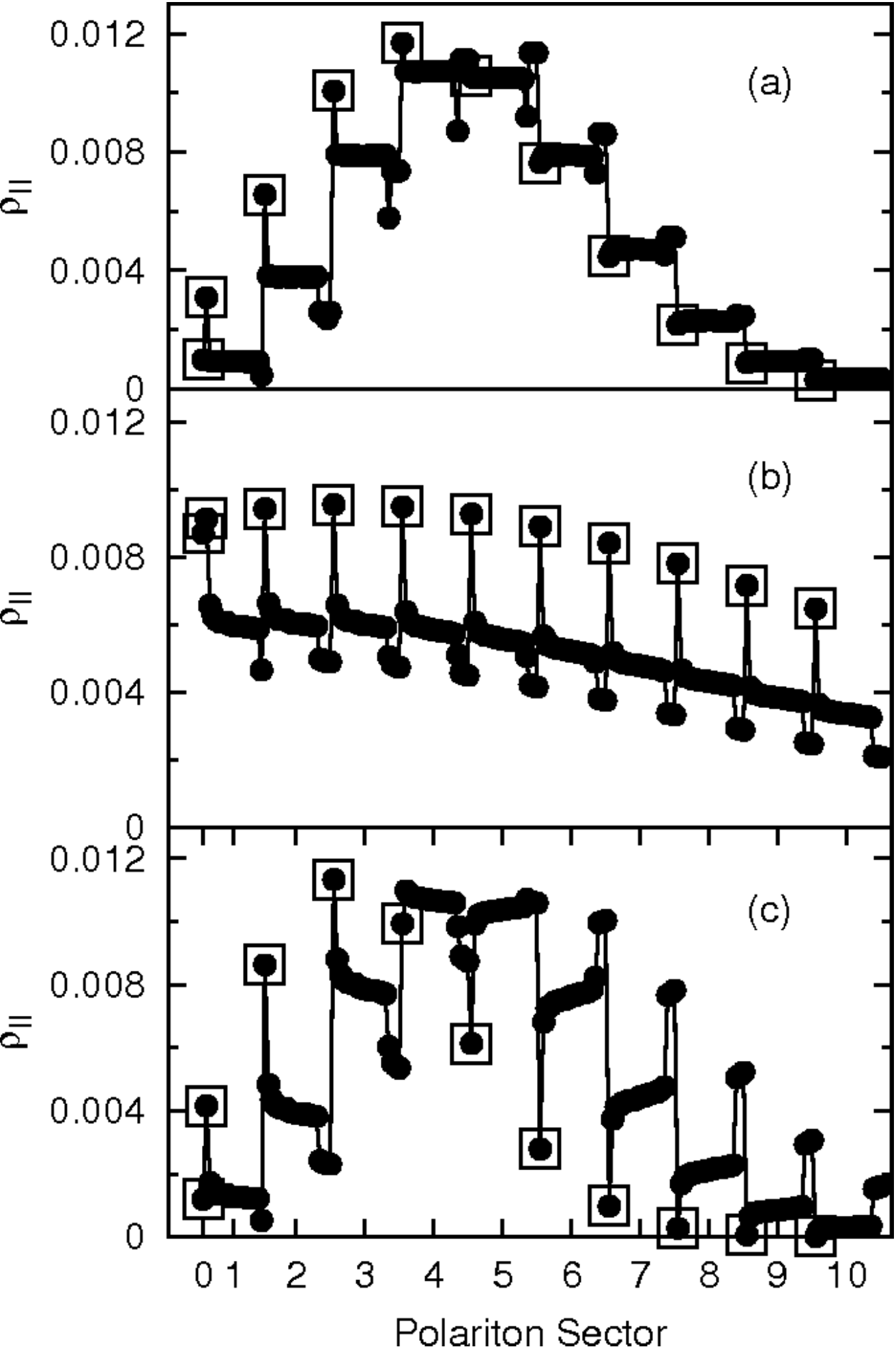}
\caption{\label{fig3} (a) The occupations in each polariton 
sector. The detuning parameter is $\Delta=-3$ meV, and the pumping 
rate is $P=0.01$ ps$^{-1}$. Ground-state occupations are indicated by 
squares. (b) The best fit to the actual occupations with a grand 
canonical Gibbs distribution. (c) Best fit with a distorted
grand canonical Gibbs distribution.}
\end{center}
\end{figure}

In the stationary limit, the r.h.s. of Eq. (\ref{eq5}) is equal to 
zero. This set of homogeneous linear equations can be shown to be
linearly dependent. Indeed, it can be easily verified that

\begin{equation}
\frac{{\rm d}}{{\rm d}t}\sum_I \rho_{II}=0,
\end{equation}

\noindent
which corresponds to the conservation of probability, 
$\sum_I \rho_{II}=1$. We replace the equation for $\rho_{11}$ by the
constraint $\sum_I \rho_{II}=1$ in order to obtain an inhomogeneous
system:

\begin{equation}
\sum_{J,K} M_{FI,JK}~\rho_{JK}=B_{FI},
\label{eq7}
\end{equation}

\noindent
where $M_{11,JJ}=1$, the rest of the matrix elements, $M_{FI,JK}$,
are obtained from Eq. (\ref{eq5}), and the components of the vector
$B_{FI}$ are zero, with the exception of $B_{11}=1$.
Only the occupations, $\rho_{II}$, and the ``horizontal'' coherences,
$\rho_{FI}$, where $F$ and $I$ are states in the same $N_{pol}$ sector,
acquire nontrivial values as a result of solving Eq. (\ref{eq7}). The
other coherences are equal to zero.

We show in Fig. \ref{fig3}(a) the resulting occupations for 
$\Delta=-3$ meV and $P=0.01$ ps$^{-1}$. The mean number of polaritons 
is $\langle N_{pol}\rangle=4.77$. In this case, the system is under the 
``polariton laser'' regime \cite{JBloch}. That is, above threshold 
($\langle N_{pol}\rangle\sim 1$) and below saturation due to Fermi
statistics ($\langle N_{pol}\rangle$ much greater than 10 in the 
present model, which is called ``photon laser'' regime in Ref. 
\onlinecite{JBloch}). 

The first point in the figure, enclosed by a square, corresponds to 
vacuum's occupation. The next 17 points corresponds to states with 
$N_{pol}=1$. Then, there are 20 states corresponding to $N_{pol}=2$, etc. 
In order to facilitate lecture of the figure, we have indicated the first 
state in each sector by enclosing it with a square. We may notice that the
ground-state occupations in sectors with $N_{pol}<\langle N_{pol}\rangle$ are
notably enhanced as compared with the rest of the states in the same
sector. This is mainly the result of a transfer of population from
``upper-polariton'' states, following from the matrix elements 
illustrated in Fig. \ref{fig1}(b). Consequently, the ``upper-polariton'' 
occupations are depressed.

The computed horizontal coherences are three orders of magnitude 
smaller than the occupations. For $\Delta=-3$ meV and $P=0.01$ 
ps$^{-1}$, we obtained $\sum_{I<J}|\rho_{IJ}|=7.4\times 10^{-4}$,
which should be compared with $\sum_I \rho_{II}=1$. This means that 
the stationary $\rho$ is approximately diagonal in the energy 
representation. This is the main result of the paper. We verified that 
the above statement holds for any values of the system's parameters.

Once shown that $H$ and $\rho$ approximately commute, we can try to
answer the question about whether the computed $\rho$ can be obtained
from a maximum entropy principle. In a first attempt, we fit the
occupations to a grand canonical Gibbs distribution, 
$\rho\sim\exp (-(E-\mu_1 N_{pol})/T_1)$. For the situation represented
in Fig. \ref{fig3}(a), we get the effective parameters, $T_1=28.05$ meV, 
and $\mu_1=E_{gap}-11.7$ meV. The r.m.s. deviation is 0.003, one 
fourth of the maximum values of $\rho$. Qualitatively, however, it 
gives a poor description of the actual occupations, as it can be seen 
in Fig. \ref{fig3}(b). 


A second possibility for the fit is motivated by the results of paper
[\onlinecite{Cabo}], where the stationary $\rho$ of a nonequilibrium
system is obtained from the maximization of the entropy with an 
additional constraint in phase space. Then, we should find the 
extremum of the functional:

\begin{eqnarray}
{\cal S}&=&-\sum_I \rho_{II} \ln  (\rho_{II}) + 
a_1 (\sum_I \rho_{II}-1)\nonumber\\
&+& a_2 (\sum_I \rho_{II} E(I)-\langle E\rangle)
+a_3 (Q(\rho)-\langle Q\rangle)\nonumber\\
&+&a_4 (\sum_I \rho_{II}N_{pol}(I)-\langle N_{pol}\rangle),
\end{eqnarray}

\noindent
where $a_1$, \dots, $a_4$ are Lagrange multipliers, and $Q(\rho)$ is the
constraint. It was argued in Ref. \onlinecite{Cabo} that the 
requirements of additivity and commutativity with the Hamiltonian
fix the form of the constraint, $Q(\rho)=\sum_I \rho_{II}^q F(I)$,
where $q$ is a kind of Tsallis index, and $F$ commutes with $\rho$. 
We choose $F=H-E_{gap}N_{pol}$. The equation for $\rho$,

\begin{eqnarray}
0&=&-\ln  (\rho_{II})-1 + a_1 + a_2 E(I)+a_4 N_{pol}(I)
\nonumber\\
&+& a_3 q \rho_{II}^{q-1} (E(I)-E_{gap}N_{pol}(I)),
\end{eqnarray}

\noindent
under the assumption that the distortion is weak, $q\sim 1$, can
be iteratively solved. Taking the first iteration, we get:

\begin{equation}
\rho\sim e^{-(E-\mu_2N_{pol})/T_2-\alpha_2 (E-E_{gap}N_{pol})
(E-\mu_2N_{pol})}.
\label{eq10}
\end{equation}

\noindent
The parameter $\alpha_2=a_3 q(q-1)$ is expected to be small. In our 
model, fitting the 
computed occupations to Eq. (\ref{eq10}), we obtain $T_2=8.42$ meV,
$\mu_2=E_{gap}-0.53$ meV, and $\alpha_2=0.0015$ meV$^{-2}$. The r.m.s.
deviation is $9\times 10^{-4}$, three times smaller than in the
previous case. Qualitatively, the obtained distribution, represented in
Fig. \ref{fig3}(c), is much more closer to the actual occupations. The
analogs of lower and upper polariton states show the highest dispersion.

\begin{table}
\begin{center}
\begin{tabular}{|c|c|c|c|}
\hline
$P$ (ps$^{-1}$) & $\mu_2-E_{gap}$ (meV) & $T_2$ (meV) & $\alpha_2$ 
 (meV$^{-2}$)\\
\hline
0.001 & -2.60 & 13.92 & 1.4 $\times 10^{-2}$ \\
0.01  & -0.51 &  8.37 & 1.5 $\times 10^{-3}$ \\
0.1   & -0.10 &  4.00 & 6.9 $\times 10^{-4}$ \\
\hline
\end{tabular}
\caption{\label{tab1} Effective parameters corresponding to the fit
given by Eq. (\ref{eq10}). $\Delta=-3$ meV.}
\end{center}
\end{table}

In Table \ref{tab1}, we give the effective parameters for pumping powers
in the interval from 0.001 to 0.1 ps$^{-1}$. In these computations, we
neglect coherences and extend Eqs. (\ref{eq7}) for the
occupations up to a maximum $N_{pol}=40$. 
We notice that the dependence of the chemical potential on
$P$ is qualitatively the same as observed in experiments \cite{Deng}. 
The effective temperature decreases with $P$, but for still higher
values of the pumping starts increasing, as in the experiments. We
stress, however, that the experimental fits cover a range in the
emission angle, whereas our $L=0$ distribution is qualitatively related 
to $k=0$ states, responsible for the emission along the normal direction.

In conclusion, we have computed the stationary density matrix of a
pumped polariton system, which comes from a dynamical master equation 
without thermalization mechanisms. The density matrix is shown to be 
approximately diagonal in the energy representation. Thus, a kind of 
quasiequilibrium is established. Although our model describes a 
quantum dot strongly interacting with the lowest photon mode of a thin 
microcavity, a quasiequilibrium of dynamical origin could also be 
present in the experiments described in Refs. 
[\onlinecite{Yamamoto,LeSiDang,otros}]. We should notice in this
respect that the similarity between lasing and a second-order
phase transition was underlined long ago in the context of laser physics
(see for example Ref. \onlinecite{libroOC}, Chapter 11, and references 
therein). The accumulated evidences on Bose-Einstein condensation of 
polaritons could be a present-day manifestation of this old idea. 

A dynamical framework could be also the basis for the computation of 
the photoluminescence spectra, second-order coherence functions, etc
in the present model \cite{laserpol}. Experimental facts such as the 
low threshold for polariton lasing, increase of linewidth with pumping 
power, etc are nicely (qualitatively) reproduced. In this context, our 
work for a multiexcitonic quantum dot \cite{PhysE35(2006)99,nuestraPRL,
laserpol} explores an intermediate (in the number of states)
region between the single-level dot \cite{Tejedor} and  
the infinite system (quantum well, see for example Refs. 
[\onlinecite{Szymanska,Quattropani}]), where the system is simple 
enough to be studied by exact diagonalization methods, but complex
enough to exhibit many of the properties of the infinite system.

We showed that the 
computed density matrix can be reasonably fitted to a weakly distorted
grand canonical Gibbs distribution. The multi-polariton
analogs of UP and LP branches are the states worst fitted.
Let us stress that the same strategy of obtaining $\rho$ from entropy 
maximization with an additional constraint in phase space can be 
applied \cite{q}, with success, to the description of experiments on 
metaequilibrium states in electron plasma columns \cite{HD}. 

This work was partially supported by the Programa Nacional de Ciencias 
Basicas (Cuba), the Universidad de Antioquia Fund for Research, and 
the Caribbean Network for Quantum Mechanics, Particles and Fields 
(ICTP).

\end{document}